\documentclass[twocolumn]{aastex631}
\usepackage{xspace}
\usepackage{tabularx}
\usepackage{multirow}
\usepackage{makecell}

\shortauthors{Jiang et al.}
\shorttitle{Dissecting the mass quenching in TNG50}

\begin{document}

\title{Dissecting the mass quenching in TNG50: Galaxy size determines
the quenching mode}

\author[0009-0006-1483-4323]{Haochen Jiang}
\affiliation{Department of Astronomy, University of Science and
Technology of China, Hefei 230026, China}
\affiliation{School of Astronomy and Space Science, University of
Science and Technology of China, Hefei 230026, China}

\author[0000-0003-1588-9394]{Enci Wang}\thanks{E-mail: ecwang16@ustc.edu.cn}
\affiliation{Department of Astronomy, University of Science and
Technology of China, Hefei 230026, China}
\affiliation{School of Astronomy and Space Science, University of
Science and Technology of China, Hefei 230026, China}

\author[0000-0002-3775-0484]{Kai Wang}\thanks{E-mail: wkcosmology@gmail.com}
\affiliation{Kavli Institute for Astronomy and Astrophysics, Peking
University, Beijing 100871, China}
\affiliation{Institute for Computational Cosmology, Department of
Physics, Durham University, South Road, Durham, DH1 3LE, UK}
\affiliation{Centre for Extragalactic Astronomy, Department of
Physics, Durham University, South Road, Durham DH1 3LE, UK}

\author[0009-0006-7343-8013]{Chengyu Ma} 
\affiliation{Department of Astronomy, University of Science and
Technology of China, Hefei 230026, China}
\affiliation{School of Astronomy and Space Science, University of
Science and Technology of China, Hefei 230026, China}

\author[0000-0002-7660-2273]{Xu Kong}\thanks{E-mail: xkong@ustc.edu.cn}
\affiliation{Department of Astronomy, University of Science and
Technology of China, Hefei 230026, China}
\affiliation{School of Astronomy and Space Science, University of
Science and Technology of China, Hefei 230026, China}
\affiliation{Institute of Deep Space Sciences, Deep Space Exploration
Laboratory, Hefei 230026, China}

\begin{abstract}
  The diminishing of star formation is accompanied by size differentiating,
  as quiescent galaxies are more compact than star-forming galaxies at fixed
  stellar mass. In order to understand how galaxy quenching is related to
  galaxy sizes, we performed a demographic study of 46 massive quiescent
  central galaxies with stellar mass from $10^{10.5}\rm M_\odot$ to
  $10^{11}\rm M_\odot$ in the TNG50 simulation. We found that, in addition to
  the triggering active galactic nucleus (AGN) feedback, galaxy size is also
  a major determinant of the quenching process, as small and compact galaxies
  are immediately quenched by the kinetic AGN feedback, while galaxies with
  large sizes are still active until strangulated by the cutoff of new gas
  replenishment. Further spatially resolved inspection reveals that this
  short and intense kinetic AGN feedback can only suppress the star formation
  within 1-2 kpc, resulting in this size-dependent effect of quenching. We also
  identify a long-term effect of a few Gyr timescale that the gas inflow rate
  is progressively suppressed after triggering kinetic feedback, which
  appears to effectively quench large galaxies entirely. We conclude that
  kinetic AGN feedback has two key roles in quenching: a short-term, intense
  effect that quenches the central 2 kpc region, and a long-term effect that
  suppresses the gas inflow rate and further quenches the entire galaxy.
\end{abstract}

\keywords{cosmological simulation; galaxy evolution; quenching of
star formation; black hole feedback}

\section{Introduction} \label{Sec.1}

The distribution of galaxies on the stellar mass-star formation rate (SFR)
plane exhibits a clear bimodal feature: a blue, star-forming main sequence and
a red, quiescent cloud, bridged by an under-populated green valley
\citep[e.g.][]{stratevaColorSeparationGalaxy2001,
baldryGalaxyBimodalityStellar2006, pengMASSENVIRONMENTDRIVERS2010}. Since
galaxies should be star-forming initially to accumulate their stellar mass,
there must be some galaxies that have undergone a transition from star-forming
to quiescent during evolution, which process is referred to as galaxy
quenching. Moreover, the bimodal distribution in the stellar population
properties are usually correlated to the bimodality in structural and kinematic
properties \citep[e.g.][]{kauffmannStellarMassesStar2003,
  driverMillenniumGalaxyCatalogue2006,
  blantonPhysicalPropertiesEnvironments2009, Hong2023,
wangKinematicBimodalityEfficient2023, Jia-24}, and this correlation implies
either direct or indirect causal links. In fact, many galaxy quenching models
are motivated by this correlation, such as morphological quenching and
angular momentum quenching. Nevertheless, the underlying quenching mechanisms
are still under debate.

Statistical analysis of observational data revealed that galaxies are more
likely to be quenched when they have high stellar mass or live in
dense regions, and these two trends are separable, which are referred to as
mass quenching and environmental quenching, respectively
\citep[e.g.][]{baldryGalaxyBimodalityStellar2006,
pengMASSENVIRONMENTDRIVERS2010, Wang-2018, Wang-18}. Subsequent studies found
that central galaxies, which live in the center of dark matter halos, are
primarily subject to mass quenching, while satellite galaxies are also affected
by the environment \citep{pengMASSENVIRONMENTDRIVERS2012}. Therefore,
comprehending mass quenching for central galaxies is not only critical to
deepen our understanding of galaxy evolution, but also necessary to isolate the
environmental effect for us to study the environmental quenching process.

One way to approach the cause of mass quenching in observation is to find other
galaxy properties that have the strongest correlation, thus are the most likely
to be causally linked to the star formation status of central galaxies. Several
such galaxy properties were indeed found, in both local and high-$z$ Universe,
and they are the bulge mass, the stellar velocity dispersion in galaxy center,
and the surface stellar mass density within 1kpc of the galaxy center
\citep[e.g.][]{cheungDependenceQuenchingInner2012,
  wakeRevealingVelocityDispersion2012, fangLinkStarFormation2013,
  wooTwoConditionsGalaxy2015, whitakerPredictingQuiescenceDependence2017,
  Wang-18b, xuCriticalStellarCentral2021a, bluckQuenchingGalaxiesBulges2022,
piotrowskaQuenchingStarFormation2022}. All these properties are linked to the
super-massive black hole (SMBH) living in the center of galaxies
\citep{kormendyCoevolutionNotSupermassive2013}. These observational results
motivate theoretical models to include the feedback from accreting SMBH, which
is referred to as the AGN feedback, and lets them quench the star formation
activity in massive galaxies \citep[e.g.][]{bowerBreakingHierarchyGalaxy2006,
  crotonManyLivesActive2006, weinbergerSimulatingGalaxyFormation2017,
schayeEAGLEProjectSimulating2015}. However, direct evidence of AGN feedback
quenching galaxies, or what counts as the direct evidence, is still under
pursuing and debating \citep[e.g.][]{fabianObservationalEvidenceAGN2012,
Wang-19, shangguanAGNFeedbackStar2020, wangtao2023}.

Regarding the difficulty in finding direct evidence of AGN feedback quenching
and depicting the detailed mechanism of AGN feedback quenching, we resort to
hydrodynamical simulations, starting from understanding how AGN feedback
operates and how it relates to the quenching of massive central galaxies,
aiming to find the unique observational signature. In this work, we carry out a
detailed study of quiescent massive central galaxies in the IllustrisTNG
simulation \citep{springelFirstResultsIllustrisTNG2018}. We find that AGN
kinetic feedback is a necessary but not always sufficient condition for
quenching. Our results indicate that for feedback to be effective, the stellar
distribution of a galaxy must be confined within a region of 1–2 kpc, which is
comparable to the effective coupling scale of the kinetic-mode AGN feedback in
our sample.

This paper is structured as follows. \S\,\ref{Sec.2} introduces the
IllustrisTNG simulation and the galaxy sample we constructed to study the star
formation quenching process. In \S\,\ref{Sec.3}, we investigate the evolution
of galaxy properties specifically before and after the triggering of kinetic
feedback, through which we identify two consequences of kinetic feedback: a
short-lived and intense feedback mainly acting on the central 1-2 kpc of
galaxies, and a long-lived and gentle feedback that persistently suppresses gas
inflow. The results and findings are summarized in \S\,\ref{Sec.5}.

\section{Data}\label{Sec.2}

\begin{deluxetable*}{ccc}
  \tablecaption{Descriptions of parameters used in this work.}
  \label{Table 1}
  \tablehead{\colhead{Parameter} & \colhead{Field} & \colhead{Description}}
  \startdata
  $\mathrm{SFR_{2hmr}}$       & {\tt SubhaloSFRinRad}
  & \makecell{Sum of the individual star formation rates\\ of gas
  cells within twice the stellar half mass radius in this subhalo}
  \\
  \hline
  $\mathrm{SFR_{  tot}}$      & {\tt StarFormationRate}
  & \makecell{Sum of the individual star formation rates\\ of all gas
  cells in this subhalo}
  \\
  \hline
  Stellar/Gas/Black Hole Mass & {\tt SubhaloMassInRadType[4]/[0]/[5]}
  & \makecell{Sum of masses of all particles/cells (split by
  type)\\ within twice the stellar half mass radius}
  \\
  \hline
  Halo Mass                   & {\tt Group\_M\_Crit200}
  & \makecell{Total Mass of this group enclosed in a sphere\\ whose
  mean density is 200 times the critical density of the Universe}
  \\
  \hline
  Size                        & {\tt SubhaloHalfmassRadType[4]}
  & \makecell{Comoving radius containing half of the stellar mass of
  this Subhalo}
  \\
  \hline
  Metallicity                 & {\tt SubhaloGasMetalFractions}
  & \makecell{The dimensionless ratio of the total mass in that
    species\\ divided by the total gas mass,\\ both restricted to gas
  cells within twice the stellar half mass radius} \\
  \hline
  $\mathrm{F_{\epsilon>0.7}}$ & {\tt CircAbove07Frac}
  & \makecell{The fractional mass of stars with $\epsilon
    \textgreater$ 0.7,\\ which is a common definition of the "disk"
  stars}                                                 \\
  \enddata
\end{deluxetable*}

\begin{figure*}[htb]
  \centering
  \includegraphics[width=\linewidth]{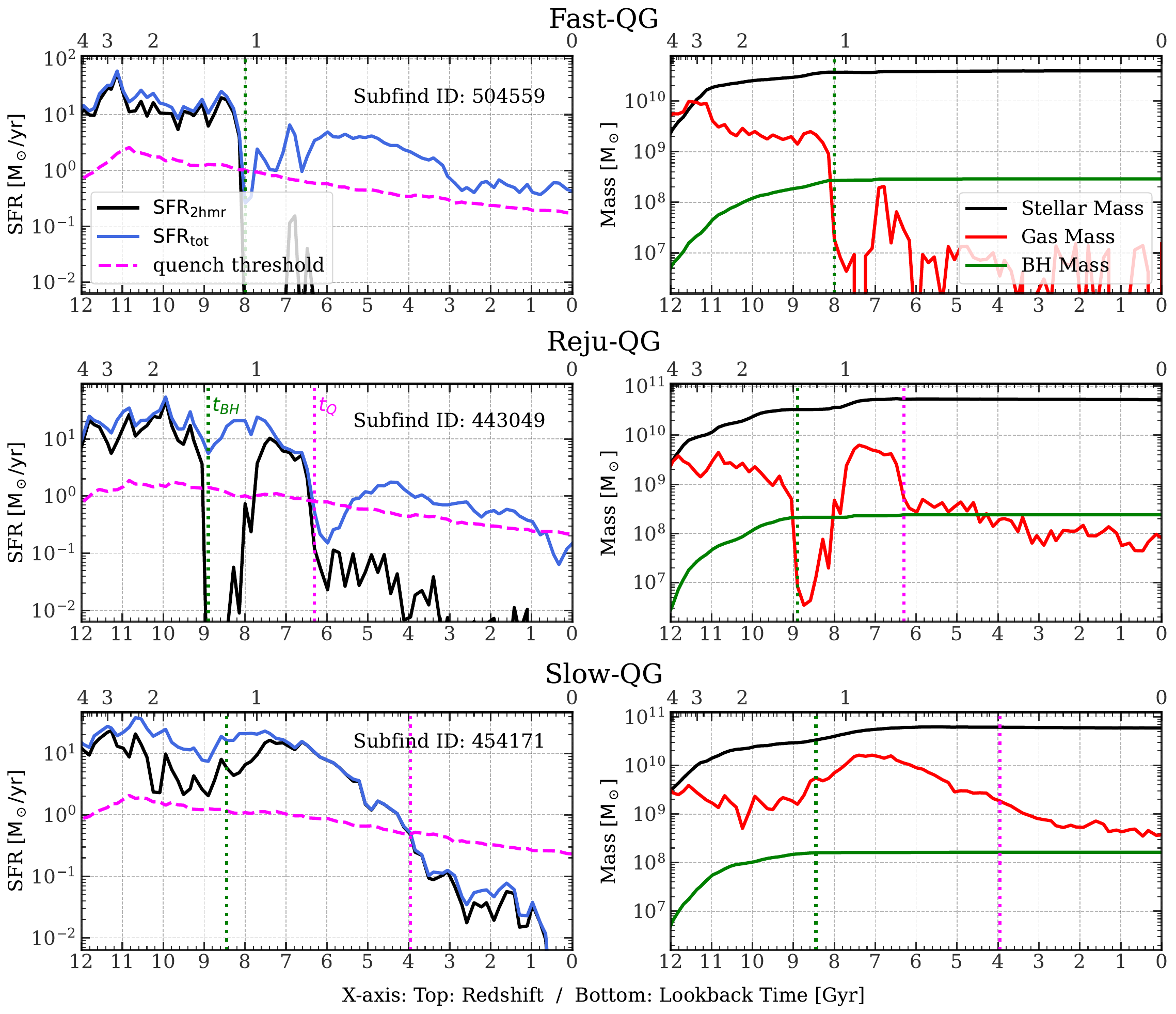}
  \caption{
    The evolution history of three example quiescent galaxies selected at
    $z=0$ from the TNG50 simulation. The left panels show the SFR evolution
    within two radii (black: twice the effective radius; blue: within the
    bound subhalo), and the magenta dashed line shows the threshold below
    which the galaxy is classified as quiescent. The lookback time and
    redshift are labeled on the bottom and top x-axis, respectively. The
    right panel shows the mass evolution for the stellar component (black),
    gaseous component (red), and the SMBH (green), defined with twice the
    effective radius. The green vertical dotted lines mark the moment the
    SMBH growth flattens, and the kinetic AGN feedback is triggered. The
    magenta vertical dotted lines mark the last quenching event. These
    three example galaxies are selected from three types of quenching
    patterns: fast-quenching ({\tt Fast-QG}), rejuvenated-quenching
    ({\tt Reju-QG}),
    and slow-quenching (\tt{Slow-QG}).
  }
  \label{Fig.1}
\end{figure*}

\subsection{The IllustrisTNG simulation}\label{2.1}

The IllustrisTNG (TNG for short) project comprises several state-of-the-art
cosmological magnetohydrodynamical galaxy formation simulations
\citep{springelFirstResultsIllustrisTNG2018,
  pillepichFirstResultsIllustrisTNG2018,
  pillepichSimulatingGalaxyFormation2018, naimanFirstResultsIllustrisTNG2018,
  marinacciFirstResultsIllustrisTNG2018, Pillepich2019, Nelson2019,
Nelson2021}, built upon the moving-mesh code AREPO \citep{Springel-10,
weinbergerAREPOPublicCode2020}. This code was implemented carefully
designed subgrid recipes for star formation, stellar feedback, the seeding
and the growth of SMBH, and the AGN feedback process
\citep{weinbergerSimulatingGalaxyFormation2017,
pillepichSimulatingGalaxyFormation2018}. Moreover, the TNG simulations can
reproduce many observational statistics
\citep[e.g.][]{nelsonFirstResultsIllustrisTNG2018, shiColdGasMassive2022,
wangDissectTwohaloGalactic2023}, including the stellar mass
functions of galaxies up to $z\sim4$
\citep{pillepichFirstResultsIllustrisTNG2018},
the color-magnitude diagram of blue and red galaxies
\citep{nelsonFirstResultsIllustrisTNG2018}, the spatial clustering of red
and blue galaxies up to tens of Mpc
\citep{springelFirstResultsIllustrisTNG2018} and etc. These, and the
publicly available output at the particle level, make the TNG one of
the best choices to study the quenching process in cosmological simulations.

The TNG simulations seed an $8\times 10^5h^{-1}\ \mathrm{M}_\odot$
super-massive black hole at the halo center, fixing it there with an ad hoc
centering prescription whenever the Friends-of-Friends halo mass exceeds
$5\times 10^{10}h^{-1}\ \mathrm{M}_\odot$ and has no black hole. Once seeded,
SMBHs grow by accreting surrounding gas at a pure Bondi rate, limited by the
Eddington rate. The feedback from accreting SMBHs is implemented in two modes.
When the accretion rate is high, the ``quasar mode'' is triggered, and the
feedback energy is thermally dumped into the surrounding medium, increasing its
temperature in a kernel fashion. However, due to the numerical overcooling
problem, this feedback energy is quickly radiated away in the presence of dense
gas around SMBHs, making it inefficient at suppressing star formation activity
\citep{katzCosmologicalSimulationsTreeSPH1996,
schayeEAGLEProjectSimulating2015}. The primary purpose of this feedback mode is
to enable SMBH growth. When the accretion rate is low, the ``radio mode'' is
activated, and the feedback energy is injected into the surrounding gas
kinetically, accelerating its velocity. In this mode, rather than distributing
energy smoothly across nearby gas cells, a minimum threshold of kinetic energy
is deposited per cell, ensuring strong coupling to the surrounding gas.
Importantly, the switch between quasar and kinetic mode is not solely
determined by the BH accretion rate but also depends on the Eddington ratio and
BH mass. Specifically, radio-mode feedback is activated when the Eddington
ratio is below 0.01 and the BH mass exceeds $10^8\,\mathrm{M}_\odot$.

The TNG simulation first identifies dark matter halos using the
Friends-of-Friends algorithm with a linking length of 0.2 using only dark
matter particles. Then, subhalos are identified using the \texttt{SUBFIND}
algorithm \citep{springelPopulatingClusterGalaxies2001} using all types of
particles, and the baryonic part of each subhalo composes a galaxy. The subhalo
with the minimal gravitational potential in each dark matter halo is defined to
be the central subhalo, and the galaxy corresponding to it is the central
galaxy. Subhalos are connected across snapshots using the merger tree
construction algorithm in \citet{springelSimulationsFormationEvolution2005},
and the main branch is constructed by recursively selecting the progenitor
subhalo with the most massive history
\citep{deluciaHierarchicalFormationBrightest2007}.

For each identified galaxy, the galaxy center is chosen to be the position with
the minimal gravitational potential, and the effective radius is measured to
enclose half of the total stellar mass. Following the recommendation of
\citet{pillepichFirstResultsIllustrisTNG2018}, we use the galaxy stellar mass
as the total mass of stellar particles within twice the effective radius, so
for the gas mass, SMBH mass, and the star formation rate. In addition,
when studying the quenching of entire galaxies, we also adopt the total gas
mass and total SFR in the analysis. The quantities used in this work are listed
in Table \ref{Table 1}.

In this work, we use TNG50, the highest-resolution version, to avoid numerical
effects in the center of these galaxies, and the side length of the simulation
box is about $50\rm \ Mpc$. The average mass for gas cells and stellar
particles is about $8.5\times 10^4\,\rm M_\odot$, and the mass of dark matter
particle is about $4.5\times 10^5\,\rm M_\odot$, which is sufficient for our
study.

\subsection{Sample Selection}\label{2.2}

Our sample comprises 46 quiescent central galaxies with stellar mass from
$10^{10.5}\,\rm M_\odot$ to $10^{11}\,\rm M_\odot$ at $z=0$ in TNG50.
The quiescent
galaxies are those that lie 1 dex below the star-forming main sequence
\citep[SFMS;][]{brinchmann04,noeske07,daddi07,speagle14} at $z=0$, which is
calibrated using an iterative algorithm
\citep{wangDissectTwohaloGalactic2023, Ma-24}.
Specifically, we start from an initial guess of a linear function and
iteratively re-fit the slope and intercept using galaxies within 1 dex of
the main sequence until convergence. For each galaxy in our sample, we track
the history of its main progenitor with the merging tree and specifically
investigate its properties around the time of triggering kinetic
feedback and the time of star formation quenching.  In doing so, the
SFMS and the quenching
criterion are calibrated at different redshifts using galaxies in the
corresponding snapshots, so that we can make a fair judgment of the star
formation status of galaxies at different redshifts. Finally, for
each galaxy in our sample, we select reference star-forming galaxies
(SFGs) that satisfy two
criteria: it is star-forming at $z=0$, and the stellar mass of its
main progenitor is within 0.1 dex of our target galaxy at the time of triggering
kinetic feedback.

\begin{figure*}[htb]
  \centering
  \includegraphics[width=0.95\linewidth]{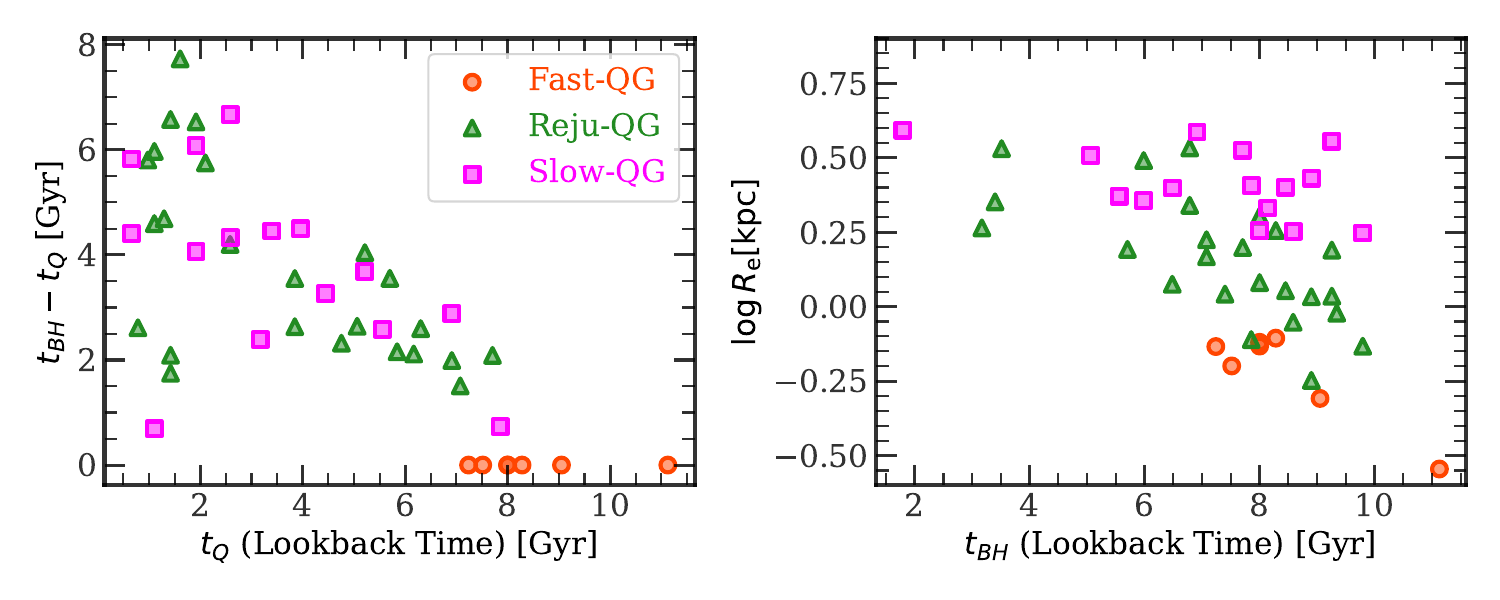}
  \caption{
    The 46 samples from the three classes (orange circle:
      \texttt{Fast-QG}; green
    triangle: \texttt{Reju-QG}; magenta square: \texttt{Slow-QG}) on
    the $t_{\rm Q}$ vs. ($t_{\rm BH} - t_{\rm Q}$) plane and the
    effective radius vs. $t_{\rm BH}$ plane. Many galaxies across all
    three classes have similar $t_{\rm BH}$ values but different
    sizes. Galaxies with smaller sizes tend to quench rapidly, while
    larger galaxies quench more slowly or incompletely.
  }
  \label{Fig.7}
\end{figure*}

\begin{figure*}[htb]
  \centering
  \includegraphics[width=0.95\linewidth]{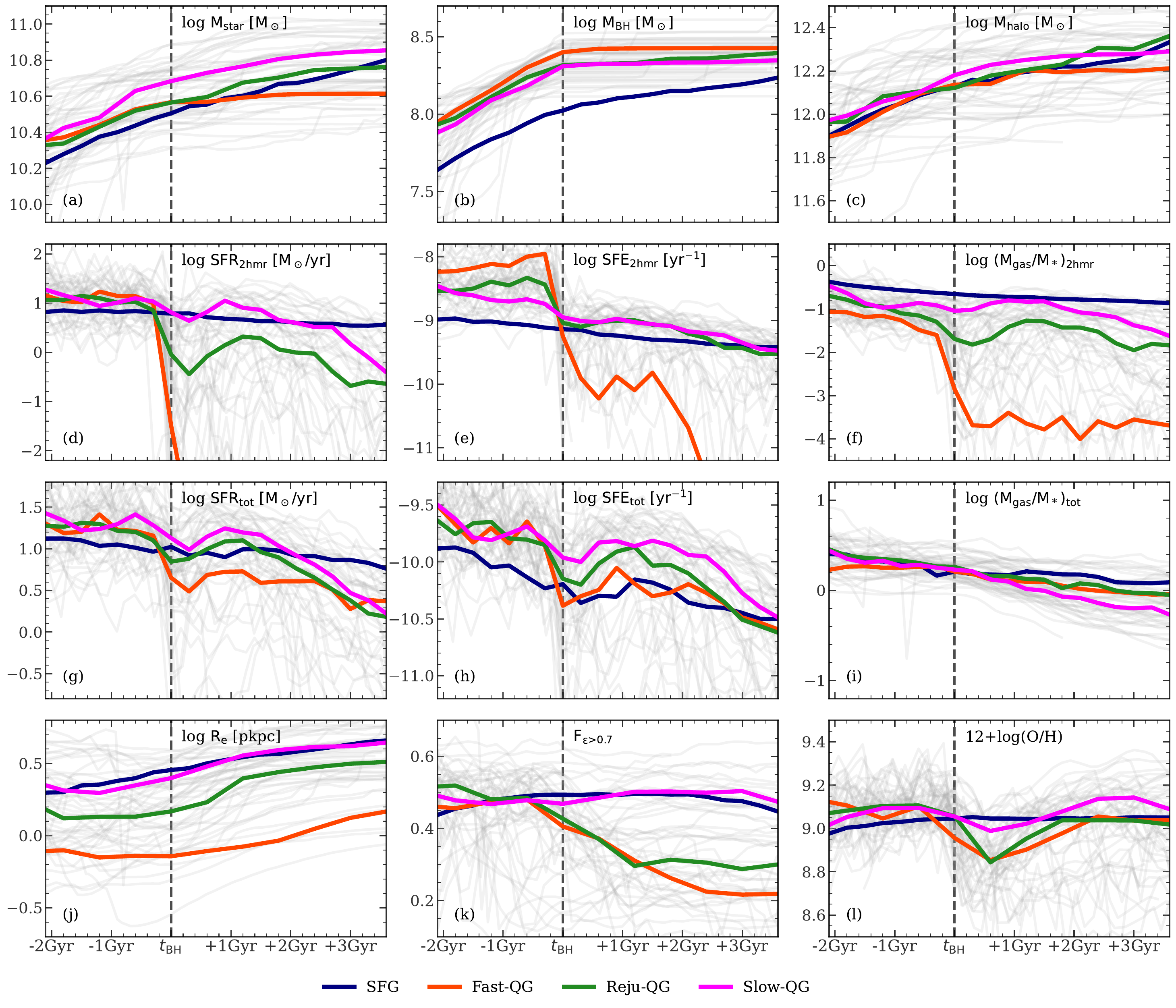}
  \caption{
    The evolution of galaxy properties around $t_{\rm BH}$ for three types
    of quiescent galaxies selected at $z=0$ (red: {\tt Fast-QG},
      green: {\tt Reju-QG},
    Blue: {\tt Slow-QG}), and the evolution of reference star-forming
    galaxies is
    shown in black. The background faded lines show the evolution
    trajectory of individual galaxies, and the solid lines show the median
    of each type of galaxies.
  }
  \label{Fig.2}
\end{figure*}

\begin{figure*}[htb]
  \centering
  \includegraphics[width=0.95\linewidth]{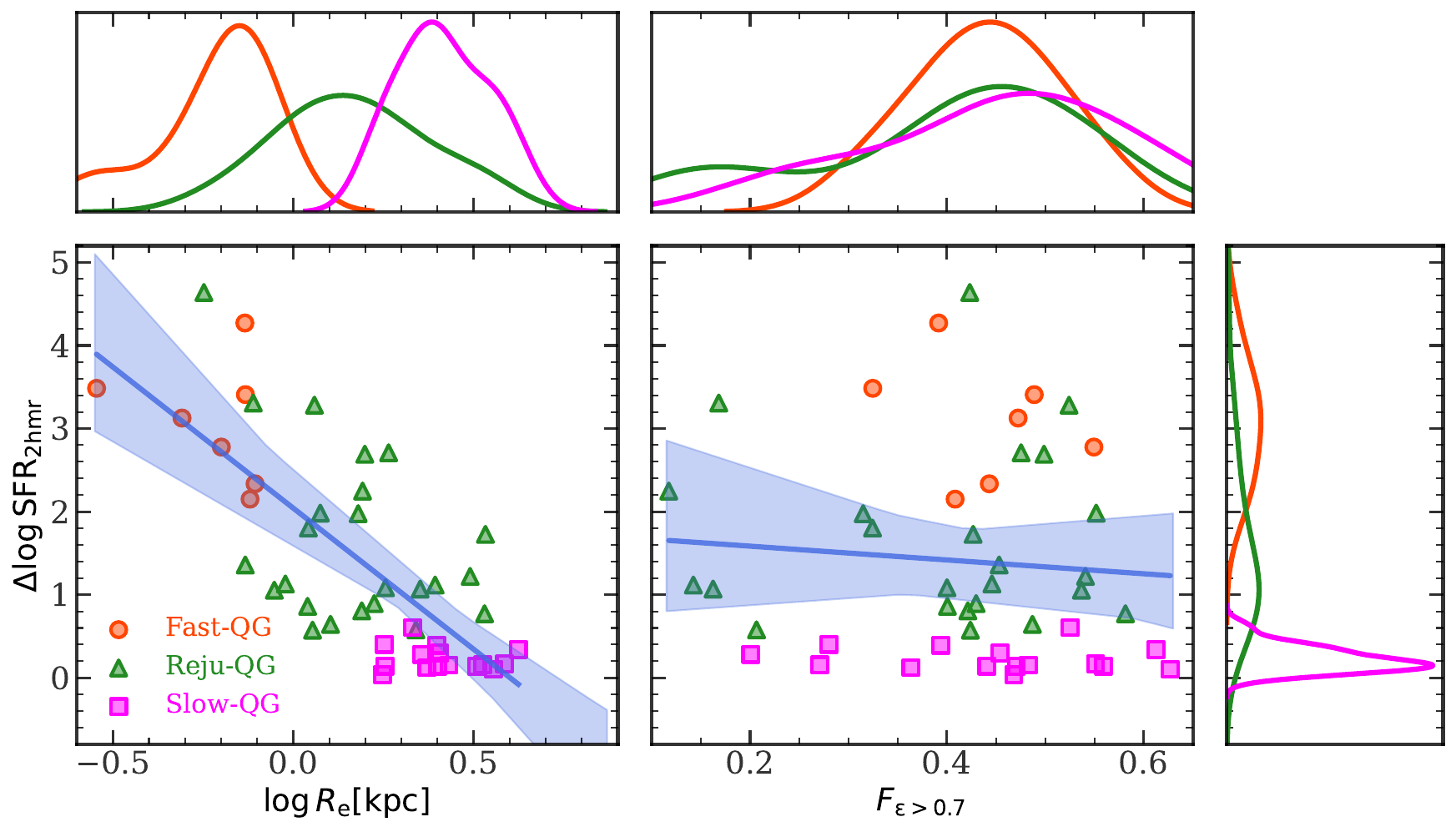}
  \caption{
    The relationship between the deviation from the star-forming main
    sequence and the effective radius (left) and kinematics (right), which
    is quantified by the fraction of stellar particles with orbital
    circularity above 0.7, measured at $t_{\rm BH}$ for three categories of
    quiescent central galaxies in TNG50.
    The blue solid lines show the linear fitting, and the shaded regions
    show the $1-\sigma$ range of bootstrap samples. The surrounding three
    small panels show the distribution of corresponding properties.
  }
  \label{Fig.3}
\end{figure*}

\subsection{Classification}\label{2.3}

Through inspecting the evolutionary histories of these quiescent galaxies, we
identified three different patterns, based on which we classify these
46 galaxies
into three types: \texttt{Fast-QG}, \texttt{Reju-QG}, and \texttt{Slow-QG}. To
begin with, this classification is built upon the definition of two critical
moments in the galaxy evolution history. The first critical moment, quenching
time $t_{\rm Q}$, is defined as the moment that the target galaxy is classified
as star-forming before and as quiescent after. Note that a galaxy may
experience several quenching times, and we only use the last one. The second
critical moment, black hole time $t_{\rm BH}$, is defined as the moment that
the kinetic AGN feedback energy begins to be injected, which can be
regarded as the
moment that the kinetic feedback is on and the subsequent SMBH growth through
accretion is marginal. For comparison, we also define a third moment
which corresponds to the third snapshot prior to $t_{\rm BH}$ for
each galaxy, denoted as $t_{\rm SF}$. This is because they are
star-forming galaxies at $t_{\rm SF}$.

Bearing these two critical moments in mind, let us inspect the evolution
histories of three galaxies in Fig.~\ref{Fig.1}, exemplifying three categories
of quiescent galaxies. In the left panels, the black and blue solid lines show
the evolution of SFR within twice the effective radius and the entire central
subhalo, respectively. The magenta dashed lines show the boundary between
star-forming and quiescent galaxies, which is time-dependent due to the
redshift-dependent SFMS and the time-varying main progenitor stellar mass. In
the right panels, the evolution of stellar mass, gas mass, and black hole mass
are shown in black, red, and green solid lines, respectively.

First of all, as shown on the top panels of Fig.~\ref{Fig.1}, galaxies
classified as \texttt{Fast-QG} have overlapped $t_{\rm BH}$ and $t_{\rm
Q}$. This means that these galaxies quench shortly after the SMBHs stop rapid
growth, and the kinetic feedback mode is on. We reason that the kinetic feedback
quickly sweeps all of the cold gas beyond twice the half-mass radius. This can
be seen from the gas mass evolution history on the top-right panel, where the
gas mass drops by more than two magnitudes and terminates the star formation
activity. Moreover, even though the star formation activity within twice the
effective radius is entirely shut down, there is still non-negligible star
formation on larger radii, which are beyond twice the effective radius.

The middle panels in Fig.~\ref{Fig.1} show one example for \texttt{Reju-QG}
galaxies, which behave similarly to \texttt{Fast-QG} initially. However, due to
the presence of additional gas inflow at a later time, they are rejuvenated for
a while. Besides, eventually, these galaxies are quenched by the depletion of
cold gas.

The last category, \texttt{Slow-QG}, is exemplified by the galaxy shown in the
bottom panels of Fig.~\ref{Fig.1}. Here, one can see that the star
formation rate,
both within twice the effective radius and the entire galaxy, slightly decreases
around $t_{\rm BH}$, but it is still above the quenching criterion. Then,
accompanied by the decreasing gas mass, the star formation activity
in this galaxy is gradually diminished until it is quenched 5 Gyr
after. The quenching process
of these galaxies is slow and gradual, in sharp contrast to the other two types
of galaxies, which are rapid and prompt.
Fig.~\ref{Fig.7} shows the $t_{\rm Q}$ vs. $(t_{\rm BH}-t_{\rm Q})$
plane of all three types of galaxies. The classification criteria can
be described as follows:

{\tt Fast-QG}: $t_{\rm Q}$ overlaps with $t_{\rm BH}$.

{\tt Reju-QG}: at $t_{\rm BH}$, the SFR falls below the quenching
threshold but rises again before final quenching at $t_{\rm Q}$.

{\tt Slow-QG}: similar to {\tt Reju-QG}, but the SFR does not fall
below the quenching threshold at $t_{\rm BH}$.

\section{Results}\label{Sec.3}

\begin{figure*}[htb]
  \centering
  \includegraphics[width=0.90\linewidth]{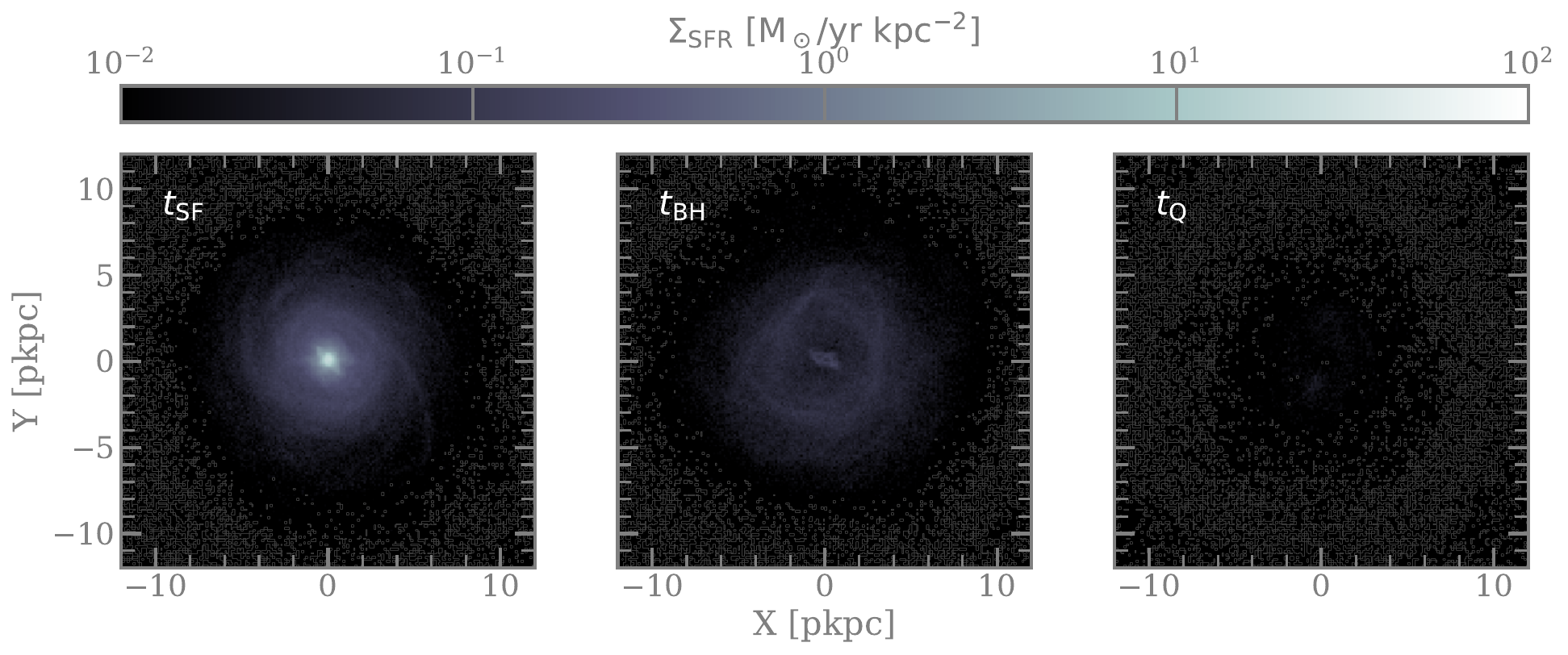}
  \includegraphics[width=0.90\linewidth]{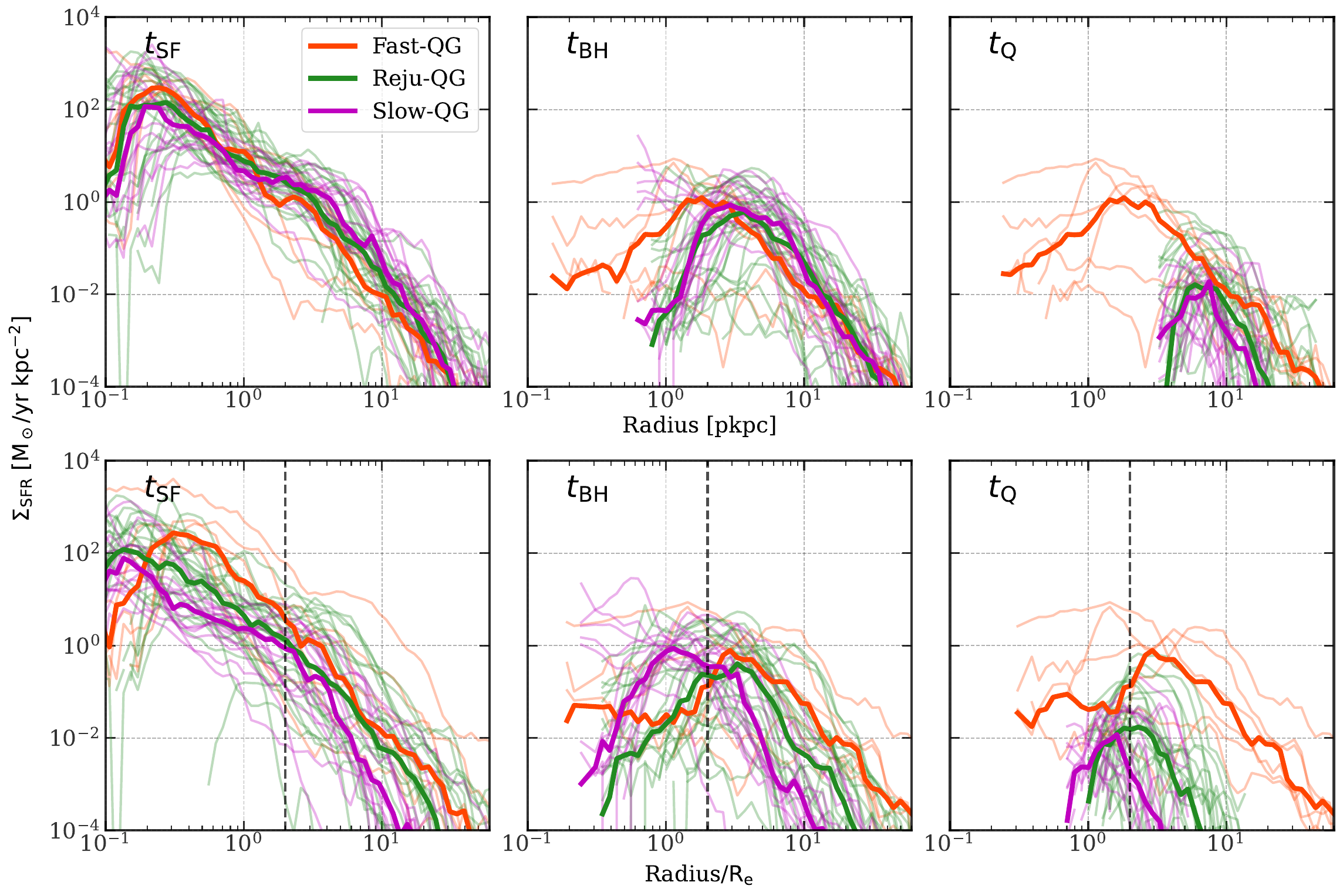}
  \caption{
    The stacked SFR surface density of three types of quiescent galaxies at
    the star-forming time ($t_{\rm SF}$, left), SMBH growth flattened time
    ($t_{\rm BH}$, middle), and quiescent time ($t_{\rm Q}$, right). The
    top panels show the stacked face-on image with a color bar encoding the
    SFR surface density. The bottom panels show the SFR surface density as
    a function of physical galaxy-centric distance and $R_{\rm
    e}$-normalized galaxy-centric distance. In each panel, the thin lines
    show the individual result, while the thick lines show the average of
    each type of quiescent galaxies. This figure clearly demonstrates that
    the triggering of kinetic AGN feedback at $t_{\rm BH}$ can effectively
    quench star formation activities within $\approx 1-2$ kpc, which causes
    those compact and small galaxies to be classified as quenched, while
    those extended galaxies are still counted as star-forming galaxies.
  }
  \label{Fig.4}
\end{figure*}

The classification of three types of central galaxy quenching motivates us to
pinpoint the driving factors of each quenching pattern, which is
critical for us to understand central galaxy quenching in the TNG
simulation. We start from
investigating the evolution of various galaxy properties around $t_{\rm BH}$,
when the SMBH growth flattens and the kinetic AGN feedback is triggered, from
which we can find clues of why these galaxies respond to the triggering
of kinetic AGN feedback differently.

\subsection{Compact galaxies are more vulnerable to kinetic AGN feedback}%
\label{3.1}

As shown in Fig.~\ref{Fig.2}, we calculate the evolution histories of three
types of massive quiescent galaxies, anchored to $t_{\rm BH}$, in red ({\tt
Fast-QG}), green ({\tt Reju-QG}) and magenta ({\tt Slow-QG}). For comparison,
we also construct a star-forming reference galaxy sample matched with stellar
mass at the time of triggering kinetic feedback.

The top three panels in Fig.~\ref{Fig.2} show the evolution of stellar mass,
SMBH mass, and the host halo mass, respectively, anchored at $t_{\rm BH}$ for
three categories of central quiescent galaxies selected at $z=0$, together with
the reference star-forming sample. At first, all three samples of \texttt{QG}
galaxies exhibit similar evolution histories of $M_{\rm BH}$, which rapidly
grows initially and flattens after $t_{\rm BH}$, signaling the triggering of
kinetic AGN feedback. This is wrote in the definition of $t_{\rm BH}$ as we
want to control the growth history of SMBH and study how star formation
activity response to the kinetic AGN feedback for galaxies with different
physical properties. Secondly, we notice that the stellar mass history of
\texttt{Fast-QG}s and \texttt{Reju-QG}s are indistinguishable to each other
around $t_{\rm BH}$; so for the halo mass history. However, we find that
\texttt{Slow-QG}s have higher stellar mass than \texttt{Fast-QG}s and
\texttt{Reju-QG}s by $\approx 0.2$ dex, and the halo mass is also higher by
$\approx 0.1$ dex, indicating the substantial difference in terms of the
quenching mechanism for \texttt{Slow-QG}s and the other two populations.

The panels on the second row of Fig.~\ref{Fig.2} present the evolution history
of SFR, star formation efficiency (SFE), and $M_{\rm gas}$, which are related
through $\text{SFR} = \text{SFE}\times M_{\rm gas}$, around $t_{\rm BH}$. Here
one can see that the SFR for both \texttt{Fast-QG}s and \texttt{Reju-QG}s
declines dramatically at $t_{\rm BH}$ by $\gtrsim 2$ dex in response to the
triggering of kinetic AGN feedback. By decomposing SFR into SFE and $M_{\rm
gas}$, we find that both factors contribute to the decline of SFR, which
indicates that the triggering of kinetic AGN feedback not only removes gas out
of galaxies, but also undermines the efficiency of making new stars. Meanwhile,
the SFR of \texttt{Slow-QG}s barely changes around $t_{\rm BH}$ and gradually
declines $\approx$ 2 Gyr after $t_{\rm BH}$, and the declination of SFR is
driven by the depletion of gas on a $\sim$ Gyr timescale instead of the
suppression of SFE. Finally, we note that all three kinds of quiescent galaxies
have higher SFE than the star-forming reference sample by $0.3-1$ dex prior to
$t_{\rm BH}$ \citep[see also][]{2025arXiv251002573W}.

The panels on the third row show similar things to the second row, except that
all quantities are measured using all bound particles for the central subhalo
instead of within $2R_{\rm e}$. According to the \texttt{SUBFIND} algorithm,
the central subhalo comprises all particles that are gravitationally bound to
the central subhalo, except those that have been associated with satellite
subhalos. Therefore, we expect all diffuse particles in a FoF halo to be
included in the central subhalo so long as the halo is in equilibrium.

Here, one can see that only \texttt{Fast-QG}s experience a dramatic change in
SFR by $\gtrsim 1$ dex, while the other two populations do not respond strongly
to the triggering of kinetic AGN feedback. This indicates that, although the
star formation activity within $2R_{\rm e}$ is quenched for \texttt{Reju-QG}s,
there are still significant star formation activities on the outskirts of
galaxies, suggesting that the kinetic feedback can only affect the inner part
of galaxies, while the outskirts stay untouched. This effect is less
significant for \texttt{Slow-QG}s and \texttt{Fast-QG}s. Last but not least,
the SFR for all three kinds of galaxies drops significantly several Gyrs after
$t_{\rm BH}$, showing that the outskirt will also be quenched given sufficient
time.

The bottom left panel in Fig.~\ref{Fig.2} shows the most notable difference
among all three categories of quiescent galaxies, which is the size evolution
before and after $t_{\rm BH}$. Here, one can see that \texttt{Slow-QG}s are the
largest, followed by \texttt{Reju-QG}s, and \texttt{Fast-QG}s are the most
compact. These three populations of galaxies are separated by $\approx 0.2-0.3$
dex. Moreover, the size of galaxies barely evolves within the timescale probed
here, except that \texttt{Reju-QG}s increase their size by $\approx 0.3$ dex
about 2 Gyr after $t_{\rm BH}$, associated with the rejuvenated star formation
activity.

The bottom middle panel shows the evolution of stellar kinematics quantified by
the fraction of stellar particles with a specific angular momentum $\epsilon$
above 0.7, i.e. $F_{\epsilon > 0.7}$. Firstly, all types of galaxies,
including the star-forming reference sample, have very similar
kinematic structure prior
to $t_{\rm BH}$. However, once passed $t_{\rm BH}$, those three quiescent
populations have distinct evolutionary tracks: \texttt{Fast-QG}s and
\texttt{Reju-QG}s drop their $F_{\epsilon > 0.7}$ from $\approx 45\%$ to
$\approx 20\%$ and $\approx 30\%$, respectively, while no declination in
$F_{\epsilon>0.7}$ for \texttt{Slow-QG}s is observed.

Fig.~\ref{Fig.3} shows how $\Delta \rm SFR$, which is the deviation from the
star-forming main sequence, correlates with the size and kinematics of galaxies
at $t_{\rm BH}$. Firstly, one can see that \texttt{Fast-QG}s and
\texttt{Reju-QG}s are $\approx 3$ and $\approx 1$ dex below the star-forming
main sequence, while \texttt{Slow-QG}s are still on the main sequence, even
though the kinetic AGN feedback is already triggered. Secondly, there is a
strong correlation between $\Delta \rm SFR$ and galaxy size: more compact
galaxies are more quenched and far away from the main sequence, while the most
extended galaxies are barely affected. On the contrary, the correlation between
$\Delta\rm SFR$ and $F_{\epsilon > 0.7}$, which quantifies the kinematic
structure of galaxies, is negligible. These results indicate that galaxy size
is a key factor strongly associated with the rapidity of quenching, though it
may not solely determine the outcome. In some cases, {\tt Fast-QG}s and {\tt
Slow-QG}s exhibit comparable effective radii, indicating that other parameters
also contribute.

A debate which may be raised is that Fig.~\ref{Fig.7} illustrates the
distribution of the 46 galaxies in the $t_{\rm BH}$ vs. ($t_{\rm BH}$
- $t_{\rm Q}$) plane. It is evident that {\tt Fast-QG}s tend to have
earlier $t_{\rm Q}$. However, this trend is not solely driven by
redshift effects. Many galaxies across all three classes share
similar $t_{\rm BH}$ values (see the right panel in
Fig.~\ref{Fig.7}). Their physical sizes at $t_{\rm BH}$ differ, and
galaxies with relatively smaller sizes tend to quench rapidly, while
larger systems exhibit delayed or incomplete quenching. This suggests
that galaxy size plays a critical role in setting the quenching
timescale and, consequently, the epoch of complete quenching ($t_{\rm
Q}$). Therefore, the observed trend that early-quenching galaxies are
more compact is a consequence of the physical conditions at the time
of AGN feedback onset, rather than a redshift-driven selection effect.

Finally, the bottom right panel of Fig.~\ref{Fig.2} shows the evolution of
gas-phase metallicity. Before $t_{\rm BH}$, all three populations of galaxies
have similar metal content. After the triggering of kinetic feedback, the
metallicity for \texttt{Fast-QG}s and \texttt{Reju-QG}s drops by $\approx 0.2$
dex, except that \texttt{Reju-QG}s are re-enriched by the rejuvenated star
formation process. Meanwhile, \texttt{Slow-QG}s are nearly unaffected.

\subsection{The influence region of kinetic AGN feedback}%
\label{sub:the_influence_region_of_kinetic_agn_feedback}

\begin{figure*}[htb]
  \centering
  \includegraphics[width=0.95\linewidth]{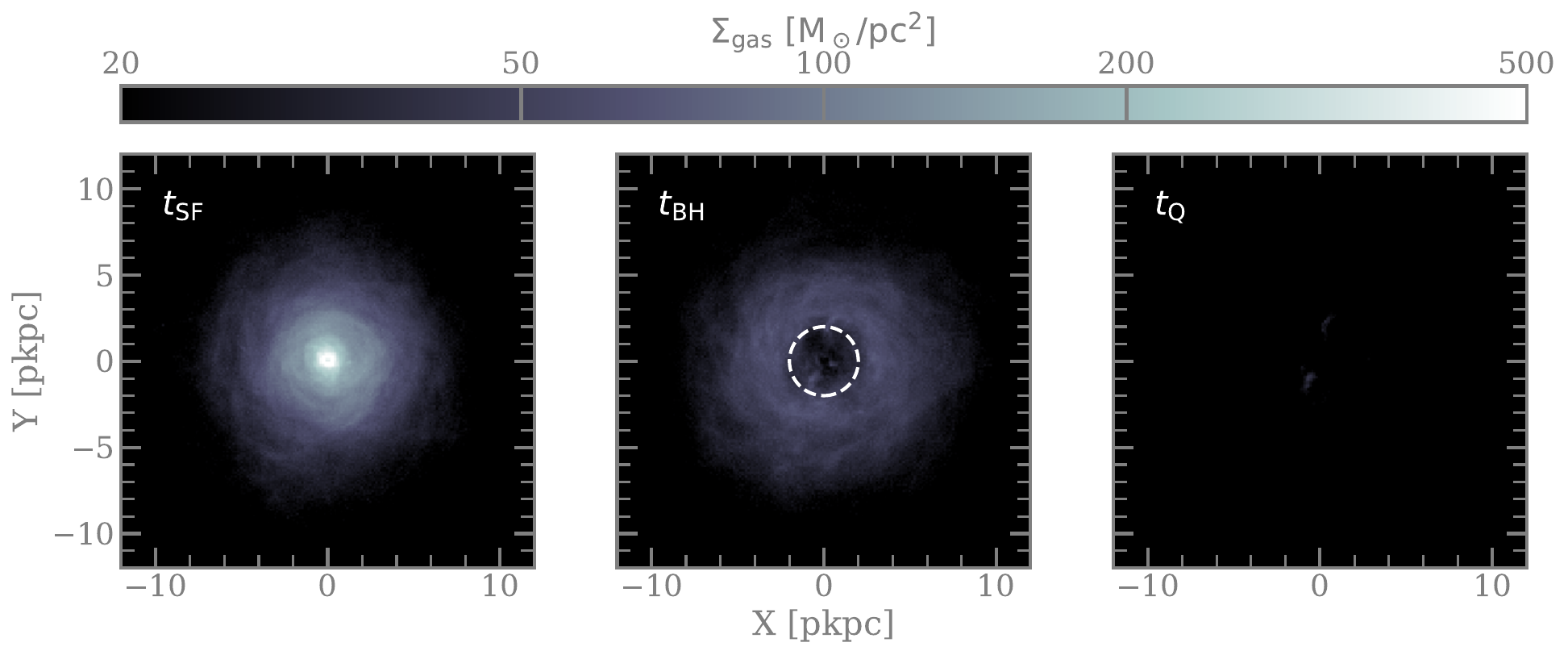}
  \includegraphics[width=0.90\linewidth]{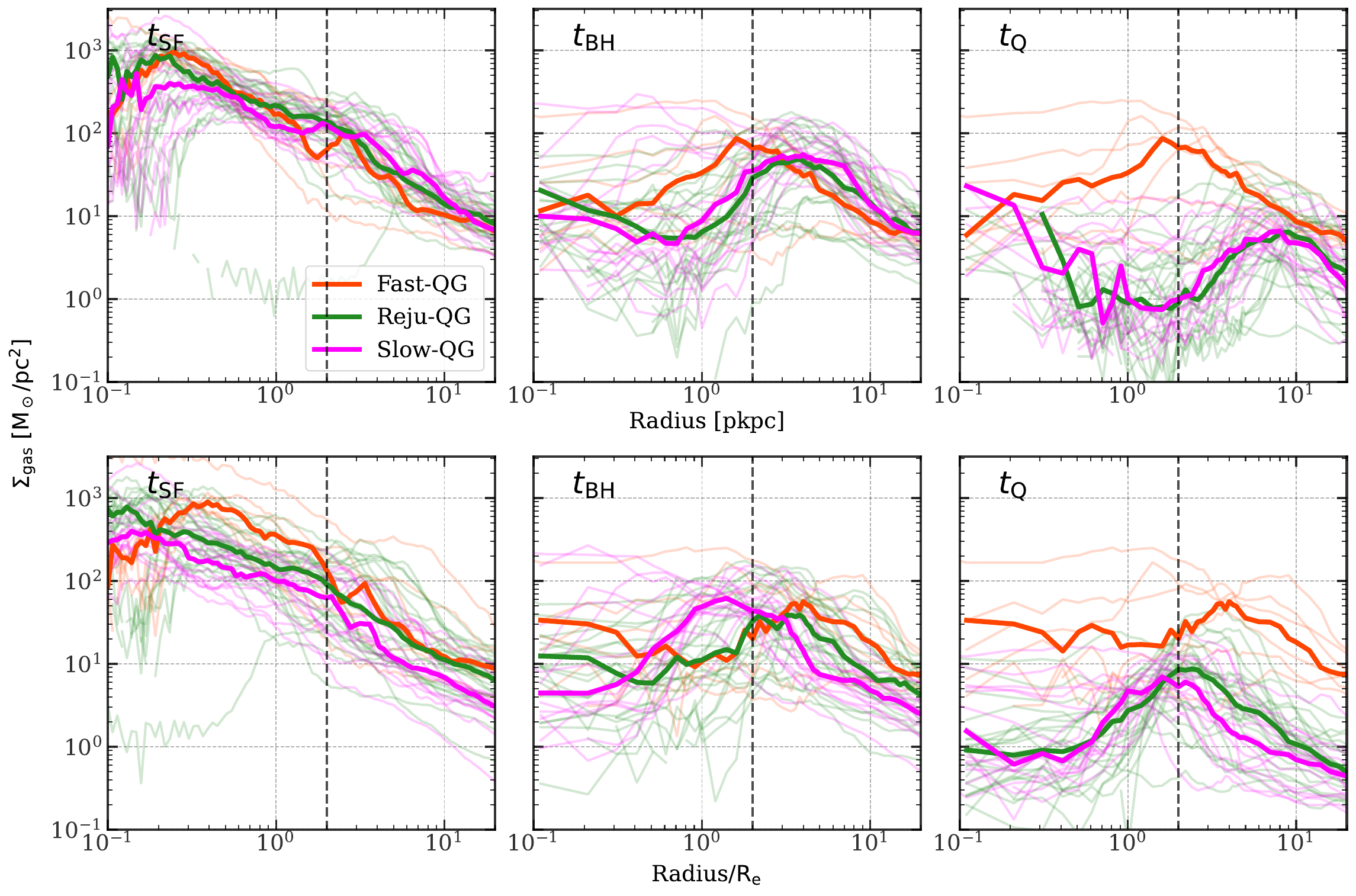}
  \caption{
    Similar to Fig.~\ref{Fig.4}, but for the gas surface density. This
    figure also demonstrates that the triggering of kinetic AGN feedback
    expels the gas content within $\approx 1-2$ physical kpc, and this
    causes those small and compact galaxies to be gas-deficient and easily
    quenched.
  }
  \label{Fig.5}
\end{figure*}

\begin{figure*}[htb]
  \centering
  \includegraphics[width=0.95\linewidth]{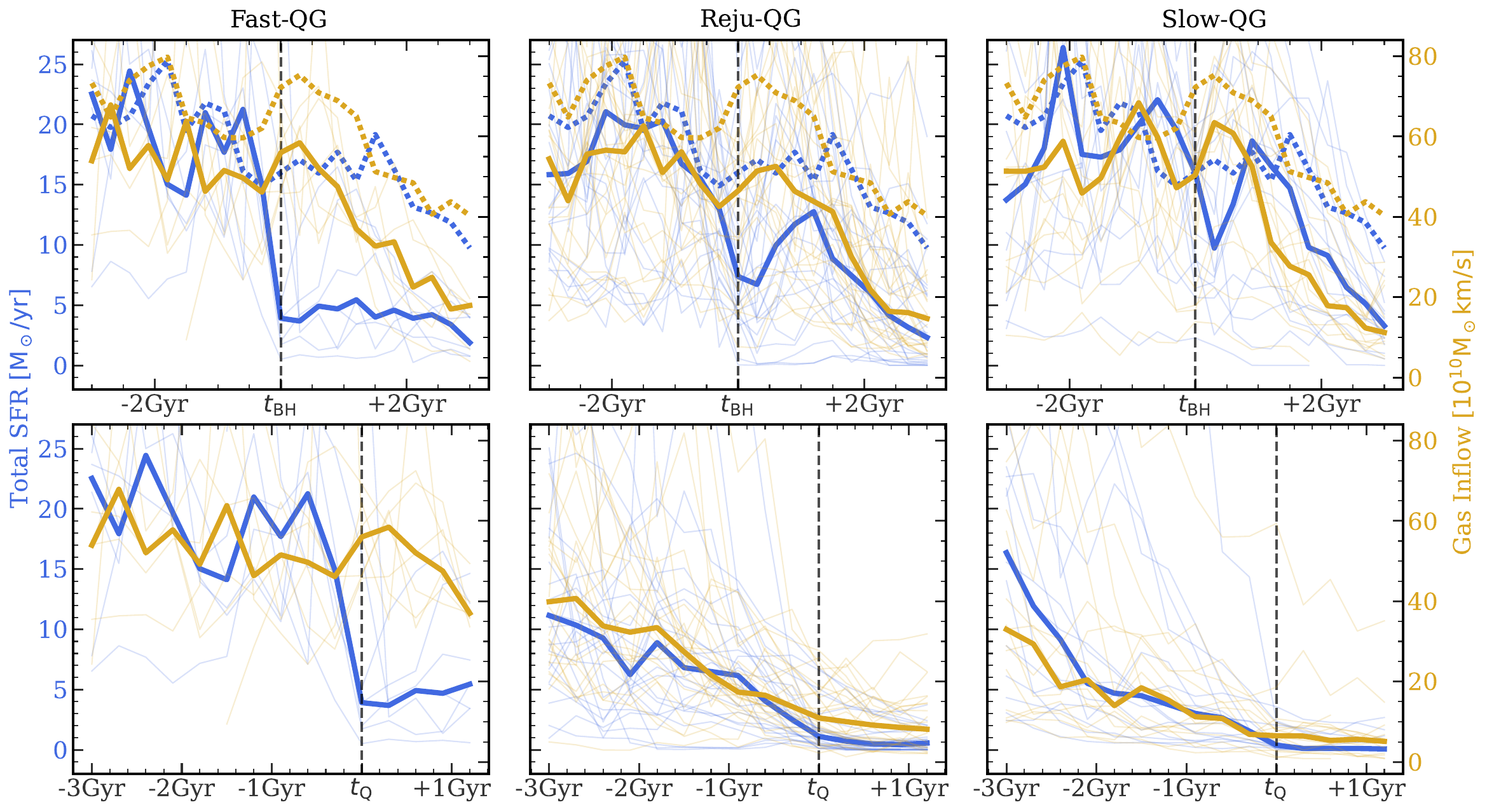}
  \caption{
    The evolution of total SFR in blue solid lines and the gas inflow rate
    in golden solid lines around $t_{\rm BH}$ and $t_{\rm Q}$ for three
    types of quiescent galaxies. The dashed lines are for the star-forming
    reference sample.
  }
  \label{Fig.6}
\end{figure*}

We have seen how rapidly central galaxies can be quenched by the triggering of
kinetic AGN feedback is determined by the galaxy size. Compact and small
galaxies are rapidly quenched after the flattening of SMBH growth and the
triggering of kinetic AGN feedback, while spatially extended galaxies are not
severely affected instantaneously, but instead slowly strangulated to the
quiescent state, accompanied by the gas consumption.

In order to figure out how the kinetic AGN feedback impacts different types
of quiescent galaxies, and how does this process relate to the galaxy size, we
plot the spatial distribution of star-forming activity and gas content in
Fig.~\ref{Fig.4} and Fig.~\ref{Fig.5}. The middle-left panels in both figures
shows that these three types of quiescent galaxies have very similar SFR
surface density ($\Sigma_{\rm SFR}$) profile and gas density ($\Sigma_{\rm
gas}$) profile when they were still actively making new stars. The
middle-middle panels further show that the triggering of kinetic AGN feedback
severely decreases the gas content, and thus diminishes the star formation
activity in the central $1-2 \rm kpc$ for all three types of quiescent
galaxies. However, when normalizing the profile using the effective radius of
individual galaxies, one can see that the star formation activity in the
central $2R_{\rm e}$ of \texttt{Fast-QG} is almost entirely quenched, simply
because their effective radius is small. Meanwhile, \texttt{Slow-QG}s can still
keep the star-forming status due to their extensive stellar mass distribution
so that the diminishing of the star formation activity of the central 1-2 kpc
cannot bring these galaxies to cross the quenching threshold. We
emphasize that this short and intense effect of kinetic AGN feedback
appears to depend only weakly on galaxy stellar surface density, as
the impact radius for all three types of quenched galaxies is
consistently around 1–2 kpc.

\subsection{Quenching by suppressing gas inflow rate}%
\label{sub:quenching_caused_by_kinetic_agn_feedback_fast_and_slow}

As we emphasized in the previous section, the short and intense effect of AGN
kinetic feedback can effectively quench galaxies in the 2 kpc region at the
center.  However, the total star formation activities of {\tt Reju-QG} and {\tt
Slow-QG} are also quenched within 3-4 Gyr after $t_{\rm BH}$, and the quenching
mechanisms stay unclear.

Here we plot the SFR and gas inflow rate evolution around $t_{\rm BH}$ and
$t_{\rm Q}$ for three types of quiescent galaxies in Fig.~\ref{Fig.6}.  The
instantaneous inflow rate is calculated as the mass flux of gas cells within a
spherical shell \citep{Nelson2019}, which is between 30 and 40 kpc from the
galaxy center. To reduce the fluctuation of the evolution curve of the
instantaneous inflow rate, Fig.~\ref{Fig.6} presents the median inflow rate
averaged over a 0.3 Gyr period, revealing a clearer evolutionary trend. For
comparison, we also include results from a constructed reference sample
comprising star-forming galaxies at redshift zero which kinetic feedback was
triggered earlier in their evolution. This comparison is useful to determine
whether the evolution of the inflow rate is caused by kinetic AGN feedback or
not. Notably, the lines for the control sample are the same in the three top
panels.

As shown in the top panels of Fig.~\ref{Fig.6}, the inflow rate of these three
types of {\tt QG} is gradually suppressed after $t_{\rm BH}$. Specifically, for
{\tt Reju-QG} and {\tt Slow-QG}, the suppression of SFR after $t_{\rm BH}$
aligns with the reduction in the inflow rate, highlighting that the cutoff of
new gas replenishment causes the final quenching of these two types of
galaxies. More importantly, the reference sample also exhibits suppressed SFR
and inflow rates following the onset of kinetic AGN feedback, although their
star formation remains unquenched even at redshift zero. This strongly suggests
that the quenching of {\tt Reju-QG} and {\tt Slow-QG} is also driven
by kinetic AGN feedback, which progressively suppresses the gas inflow rate.

We conclude that in TNG50, kinetic AGN feedback has two key roles in quenching:
a short-term, intense effect that quenches star formation within the
central 2 kpc region, and a long-term effect that suppresses the
overall inflow rate, ultimately leading to the quenching of the entire galaxy.

\section{Discussion}\label{Sec.4}

\subsection{Comparision with Other Works}\label{4.1}

As stressed in \S\,\ref{Sec.1}, the bimodality of SFR and structural parameters
indicates a physical transition from star formation to quiescence, thus galaxy
quenching, and this process is expected to be associated with
structural transformation,
either directly or indirectly. Identifying the underpinning mechanism has been
a challenging problem for decades. Recent evidence from both
observations \citep[e.g.]{wooTwoConditionsGalaxy2015,
xuCriticalStellarCentral2021a, Xu2022, Bluck2023} and simulations
\citep[e.g.][]{Weinberger2018} shows that the feedback from the
accretion of the central massive black hole plays an essential role
here. A substantial advance is the realization that it is the
cumulative AGN feedback, rather than the instantaneous feedback
power, that correlates with the quenching of central galaxies and
possibly with their morphological transformation.

Many works have studied the quenching of galaxies in TNG simulations in recent
years.  Quite a number of works, including \cite{Xu2022}, \cite{Ma2022}, and
\cite{Bluck2023}, have emphasized the importance of AGN feedback. Early TNG
simulations, such as \cite{Weinberger2018}, believed that the quenching of star
formation in galaxies occurred simultaneously with the kinetic mode, but the
examples we showed challenge this statement. Some galaxies in our sample did not
quench at the moment of black hole feedback, and most of the sample galaxies do
not completely quench even after experiencing it. From
this, we find that the size of the galaxy largely determines the quenching
characteristics of the galaxy. The larger the galaxy, the better it can
maintain star formation activity after feedback. Specifically for the slowly
quenching galaxies in this work, we see that this type of galaxy also has a
larger stellar mass and a larger proportion of disk stars. \cite{Walters2022}
investigated the reasons for diverse quenching time scales. They believe
potential well depth and gas angular momentum are decisive. Combined with the
work of \cite{Walters2022}, we believe that the larger the stellar mass and
size, and the more stable the disk structure of the galaxy, the slower the
galaxy quenching process will be.

In addition, \cite{Park2022} proposed the view that the quenching of galaxies
is unrelated to morphological transformation, and our results prove this point.
It can be considered that the quenching of galaxies and the transformation of
galaxy morphology are both the common results of other factors, namely the
black hole kinetic feedback, and there is no causal relationship between the
two. \cite{Bluck2023} also focused on the importance of black hole mass, which
can also be reflected in our work results. In Fig.~\ref{Fig.3}, the black hole
mass of the star-forming galaxy is significantly lower than that of the three
types of galaxies. This difference comes from the different evolution histories
of galaxies. The lack of galaxy merger events may have led to the slower black
hole mass growth in star-forming galaxies. This aspect has been confirmed by
\cite{Xu2022} and other works. Therefore, the black hole kinetic feedback in
the galaxy occurs later, and the galaxy already has a large enough mass and
size at this time. According to our judgment, the quenching process of the
galaxy proceeds more slowly, and it will be in the star formation stage for
a long time.

Different from other works using relatively large samples, we meticulously
study the evolution history of 46 massive central quiescent galaxies. We find
that the occurrence of black hole kinetic feedback is a necessary condition for
the quenching of massive galaxies. More importantly, we propose two effects of
kinetic feedback: a short-term one and a long-term one. \cite{Ma2022} analyzed
integrally using Illustris, TNG, and SIMBA, indicating that the black
hole plays a
dominant role in galaxy quenching. In our work, black hole kinetic feedback is
a necessary condition for the quenching of massive galaxies.

\subsection{Predictions for Observations}\label{4.2}

One of the most important purposes of using simulation to study
astrophysical processes is to provide explanations and predictions
for actual observations. The greatest advantage of cosmological
simulations is their ability to trace the evolution of galaxies and
other celestial objects over long timescales, thereby helping us
better understand the results obtained from observational studies.
In this paper, we argue that the kinetic feedback is critical for
quenching galaxies. However, it is not straightforward to directly
observe the onset of kinetic feedback, but we can infer from other
physical properties. For example, the observed higher SFE, especially
for compact galaxies, is a potential signature. Besides, blue-shifted
absorption lines represent strong outflow. Therefore, the
simultaneous existence of both is of great possibility for AGN
kinetic feedback. In addition, the suppression of star formation and
cold gas at galactic centers can potentially be a signature of
kinetic feedback. We do see some cases in MaNGA observations and HI
distribution in nearby galaxies.

These observational clues are consistent with our simulation-based
predictions about the gas dynamics during quenching. According to our
research on gas inflow, there should be significant gas inflow in the
observation of massive star-forming disk galaxies; while for massive
quenching disk galaxies or elliptical galaxies, the gas inflow should
be relatively less or completely disappear. If the mass of the
central black hole can be measured more accurately, then the result
will be that the black hole mass in the star-forming galaxy is
smaller. Recently, people have observed some quenched galaxies with
internal quenching and external star formation using integrated
spectral units. In the future, the James Webb Telescope (JWST) and
some other state-of-the-art observation equipment will obtain more
observational data at medium and high redshifts. The fast-quenching
galaxies we discussed in this work quenched at an early stage, so we
infer that as the redshift increases, the proportion of small-radius
galaxies and elliptical galaxies in the massive quenching galaxy
sample will also increase.

\section{Summary}\label{Sec.5}

How the star formation activities of massive central galaxies are quenched and
how this process is related to the sizes of galaxies stays at the heart of
galaxy evolution. In order to obtain a physical insight, we performed a
detailed demographic study of 46 central massive quiescent galaxies within
$10^{10.5} - 10^{11} \rm M_\odot$. We categorized these galaxies into three
types based on the relationship between the triggering of kinetic AGN feedback
and their quenching process. We found that a key determinant to differentiate
these three types of galaxies is their sizes. {\tt Fast-QG} has the smallest
sizes so that they can be immediately quenched by the kinetic AGN feedback.
{\tt Reju-QG} is also immediately quenched by later rejuvenated from the new
external gas supplement. {\tt Slow-QG} has the most extended stellar mass
distribution and stays star-forming after the kinetic AGN feedback for a while
until strangulated by the cutoff gas replenishment (see Figs.~\ref{Fig.2} and
\ref{Fig.3}).

The detailed spatially resolved analysis reveals the ``influence region'' of
the kinetic AGN feedback, within which the violent AGN feedback can suddenly
extinguish the star formation activity. Thus, if a galaxy has a size smaller
than the influence region, it can be immediately quenched. This explains the
quenching of {\tt Fast-QG}s, which have the smallest sizes. Otherwise, the
galaxy can still keep star-forming for a while until the gas replenishment is
cut off (see Figs.~\ref{Fig.4} and \ref{Fig.5}). Meanwhile, we indeed see that
the triggering of kinetic AGN feedback can suppress the following gas inflow on
a few Gyr timescale, which is commonly known as the preventive AGN feedback
(see Fig.~\ref{Fig.6}). Our study reveals a deep connection between galaxy
quenching and galaxy size, connected by the influence region of kinetic AGN
feedback.

TNG simulations provide several testable observational effects that could serve
as potential signatures of AGN feedback. The AGN kinetic feedback can suppress
gas within 1-2 kpc of the galactic center, leading to quenching, which can be
directly observed through atomic and/or molecular gas surveys. Additionally,
this feedback can also temporarily lower the gas-phase metallicity, opposite to
predictions from the fundamental metallicity relation \citep{Mannucci-10,
Ma-24}, creating a detectable signature of simultaneous declines in SFR and
metallicity. TNG simulations further suggest that the dynamical state of
galaxies is a result of quenching, rather than the reason, which could be
tested by examining recently quenched galaxies. Therefore, future work will be
needed to further test and validate these predictions.

\begin{acknowledgments}

  The authors thank Volker Springel for insightful comments and suggestions. KW
  acknowledges support from the Science and Technologies Facilities Council
  (STFC) through grant ST/X001075/1. EW thanks the support of the National
  Science Foundation of China (Nos. 12473008) and the Start-up Fund of the
  University of Science and Technology of China (No. KY2030000200). This work is
  supported by the National Science Foundation of China  (NSFC, Grant No.
  12233008 \& 12473008), the National Key R\&D Program of China
  (2023YFA1608100),
  the Strategic Priority Research Program of the Chinese Academy of Sciences
  (Grant No. XDB0550200), the Cyrus Chun Ying Tang Foundations, and the 111
  Project for ``Observational and Theoretical Research on Dark Matter and Dark
  Energy" (B23042).
  The IllustrisTNG simulations were undertaken with compute time awarded by the
  Gauss Centre for Supercomputing (GCS) under GCS Large-Scale Projects GCS-ILLU
  and GCS-DWAR on the GCS share of the supercomputer Hazel Hen at the High
  Performance Computing Center Stuttgart (HLRS), as well as on the machines of
  the Max Planck Computing and Data Facility (MPCDF) in Garching, Germany.

\end{acknowledgments}

\bibliography{TNG_QUENCH}{}
\bibliographystyle{aasjournal}

\end{document}